\documentstyle[aps,prd,preprint,tighten,eqsecnum,amssymb,epsf]{revtex}

\begin{document}
\draft

\title{The Asymmetric Merger of Black Holes}

\author{Sascha Husa and Jeffrey Winicour}

\address{Department of Physics and Astronomy,\\
         University of Pittsburgh}
\maketitle

\begin{abstract}
We study event horizons of non-axisymmetric black holes and show how features
found in axisymmetric studies of colliding black holes and of toroidal
black holes are non-generic and how new features emerge.
Most of the details of black hole formation and black hole merger are known
only in the axisymmetric case, in which numerical evolution has successfully
produced dynamical space-times.
The work that is presented here uses a new approach to construct
the geometry of the event horizon, not by locating it in a given
spacetime, but by direct construction.
In the axisymmetric case, our method produces the
familiar pair-of-pants structure found in previous numerical simulations of
black hole mergers, as well as event horizons that go through a toroidal epoch
as discovered in the collapse of rotating matter.
The main purpose of this paper is to show how new -- substantially different
-- features emerge in the non-axisymmetric case. In particular, we show how
black holes generically go through a toroidal phase before they become
spherical, and how this fits together with the merger of black holes.
\end{abstract}


\section{Introduction}

Not very much is known about the collision and merger of black holes in
general relativity. Previous results mainly concern the axisymmetric
head-on collision, in which numerical evolution has successfully produced
computational space-times \cite{shteuk,annin1,torus,ccode,annin2,libson}.
In recent work \cite{ndata}, the geometric details of the event horizon in such
a
collision, first obtained by locating the horizon in numerical space-times
\cite{torus,science,masso},
has been described in terms of a simple analytic model.

However, the axisymmetry of the head-on collision is a non-generic feature.
Here we analyze the event horizon for a class of generic examples of
black hole collisions and obtain features of the coalescence which differ
substantially from the axisymmetric case.

The event horizon ${\cal H}^+$ of a black hole space-time is the boundary of
the space-time region visible to observers at future null infinity ${\cal
I}^+$. Assuming that there are no singularities lying on ${\cal H}^+$, general
theorems \cite{hawkell,mtw,wald}
imply that ${\cal H}^+$ is a null hypersurface whose light rays
extend into the past until they either caustic with neighboring rays at a set
of points ${\cal C}$ or intersect a non-neighboring ray at a crossover set
${\cal X}$. Together ${\cal X}$ and ${\cal C}$ form the set of points
where light rays from outside the black hole join the horizon.  Catastrophe
theory \cite{thom,arnold,berry} shows that a caustic set consisting of a single
point
is structurally unstable, i.e. a small perturbation can produce qualitative
changes in the features. This applies to the point caustic associated with
black hole formation in the Oppenheimer-Snyder model of spherically
symmetric collapse.

The simplest structurally stable caustics are folds and cusps, which are the
only stable caustics which can occur in the axially symmetric case. These
caustics were found in the numerical studies of the head-on collision of
axisymmetric black holes \cite{science} and in gravitational collapse of an
axisymmetric
rotating cluster \cite{toroid}. An analytic solution of the curved space
horizon geometry
which reproduces these numerical results has been obtained by a conformal
transformation of the intrinsic geometry of the flat space null hypersurface
produced by the wavefront emanating from a convex, topologically spherical
surface ${\cal S}_0$, which is embedded in a Euclidean
time slice $\hat t=0$ of Minkowski space \cite{ndata}. The boundary of the past
of ${\cal
S}_0$ is generated by null geodesics normal to ${\cal S}_0$. The black hole
event horizon is modeled conformally on the flat space null hypersurface $\hat
{\cal
H}^+$ consisting of the outgoing portion of this causal boundary and its
extension to ${\cal I}^+$.

Followed into the future,
$\hat {\cal H}^+$ expands to asymptotically approach an infinite surface with
conformally spherical geometry. The conformal transformation to the curved
space
horizon ${\cal H}^+$ is tailored to stop this expansion in accord with the
final
equilibrium of a black hole. Followed into the past, the generators of $\hat
{\cal
H}^+$ trace out a smooth null hypersurface with a non-smooth boundary
consisting of
caustics ${\cal C}$ and crossover points ${\cal X}$. Again followed into the
past, the
$\hat t =const$ slices ${\cal S}_{\hat t}$ of $\hat {\cal H}^+$ shrink to zero
as they
pinch off at ${\cal C}$ and ${\cal X}$. Provided the conformal transformation
is
non-singular and single-valued on ${\cal X}$, the curved space horizon can be
consistently constructed to pinch off at the same caustic and crossover points
so that ${\cal H}^+$ and $\hat {\cal H}^+$ can be identified. This will be the
case considered here, although more general constructions are possible.

As a result of the conformal transformation, the affine parameter $\hat t$
along the flat space wavefront has to be replaced by a new affine parameter
$t(\hat t)$ for the curved space horizon.
This new affine parameter $t(\hat t)$ is chosen such that the new geometry
satisfies the single Einstein equation (the focusing equation) constraining
the intrinsic geometry of the horizon. This equation takes the form of one
ordinary differential equation (ODE) along each ray.
The values of the parameter $t(\hat t)$ along each ray are uniquely fixed by
the conditions that $t=\hat t = 0$ on ${\cal S}_0$ and
$dt/d\hat t \rightarrow 1$ as $t\rightarrow \infty$.

Because the topology of a spatial slice of a null hypersurface depends on the
particular time-slicing, the 2-dimensional leaves of the $t$ and $\hat t$
foliations have the same topology at late times, but they differ qualitatively
at early times where they intersect ${\cal C}$ and ${\cal X}$.
For the head-on collision of axisymmetric black holes, ${\cal S}_0$ is chosen
to be a prolate spheroid and it is the foliation ${\cal S}_t$ in terms of the
curved space affine parameter that gives rise to the pair-of-pants structure
of the event horizon \cite{ndata}.

Although the cusps and folds formed in the axisymmetric head-on collision are
structurally stable, there is a non-generic line on the symmetry axis whose
points are both caustics and crossover points. The horizon pinches off along
this line to give rise to the separate black holes. In the absence of
axisymmetry this crossover line would be expected to broaden into a 2-surface.
Generically, the intersection of two 2-surfaces is a curve and it is the
history of this curve, as different portions of the wavefront cross, that
corresponds to the entire 2-dimensional set of crossover points that initiate
a black hole horizon.
Note that the intersection of two null surfaces, and thus in particular
the crossover surface, is generically spacelike.
In the axisymmetric head-on collision, this curve
degenerates to a point and its history to the line that forms the trouser seam
in the pair-of-pants picture.

In the absence of axisymmetry, it has been argued \cite{siino2} that in the
formation of
a generic event horizon toroidal black holes are more probable than spherical
black holes and should be observed more frequently in an appropriate
time-slicing. This leaves some vagueness as to how toroidal topology fits in
with the formation of individual black hole and their coalescence.
The main purpose of this paper is to describe and clarify such features of
binary black hole formation and coalescence in the generic case where
${\cal S}_0$ is chosen to be a triaxial ellipsoid. We show that the individual
black holes can form with spherical topology but that there is typically
an intermediate toroidal epoch as they merge into a final spherical black hole.
(See the discussion in Sec. \ref{sec:discussion} for the relevance to
topological
censorship.)

In Section \ref{sec:flat} we
present the essential geometry of the flat space wavefront. In geometric
optics, the chief consideration in the study of wavefronts is the
2-dimensional caustic surface where classically the intensity of a beam would
be infinite. In contrast, the dominant structure in horizon formation is the
crossover set. We first describe some elementary generic properties of this
set. Then we consider as a particular model the class of wave fronts
that emanate from an ellipsoid.

In Section \ref{sec:hor} we describe the construction of the conformally
related horizon and its properties. Since the event horizon of a {\em black
hole} is defined as the boundary of the past of ${\cal I}^+$ and naturally
calculated in a backwards-in-time fashion, it is conceptually simpler to
describe the horizon structure from the time reversed point of view of a {\em
white hole} horizon (the boundary of the future of ${\cal I}^-$). In this
description, the conformally related flat space null hypersurface lies on the
ingoing (future directed) wavefront from the convex surface ${\cal S}_0$. The
white hole horizon ${\cal H}^-$ is a null hypersurface with the property that
its surface area decreases into the future and has a a finite asymptotic limit
in the past. The generating null rays leave ${\cal H}^-$ at future endpoints
consisting of caustics and crossover points. All our results are easily
restated in a time reversed sense to the black hole case.
As in Section \ref{sec:flat}, we first discuss general properties and then
specialize to the conformally ellipsoidal case.

The phenomenology found for conformally ellipsoidal horizons is presented in
Section  \ref{sec:results}, and we conclude with a discussion of our
results in Sec. \ref{sec:discussion}.

\section{Flat space horizons}
\label{sec:flat}

\subsection{General Properties}\label{subsec:flat:general properties}

The model of a white hole horizon presented in Sec. \ref{sec:hor} is based
upon the ingoing wavefront from a smooth convex surface ${\cal S}_0$ embedded
at constant time ${\hat t}=0$
in Minkowski space. The rays tracing out such a wavefront generate a smooth
null hypersurface until they reach endpoints on the boundary of the future of
${\cal
S}_0$. Let $\hat {\cal H}^-$ denote this null hypersurface, along with its
future
endpoints and its extension to past null infinity. The endpoints of the null
rays
generating $\hat {\cal H}^-$ consist of a set of caustic points ${\cal C}$,
where
neighboring rays focus, and a set of non-focal crossover points ${\cal X}$,
where
distinct null rays meet.

The generic features of our model stem from the structurally stable properties
of
the caustic-crossover set of ${\cal S}_0$. The requirement that a property be
generic is that it be preserved under arbitrary small smooth perturbations of
${\cal S}_0$. More precisely, if ${\cal S}_0$ is described in the parametric
form
$x^i=f^i(\lambda_1,\lambda_2)$ then a generic feature is preserved under smooth
deformations $f^i\rightarrow f^i+\epsilon g^i(\lambda_1,\lambda_2)$ for
sufficiently small $\epsilon$. The structurally stable caustics have been
completely classified and their local properties described
\cite{thom,arnold,berry}.
We are not aware of
an analogous treatment of the generic properties of the crossover set, which is
a
global problem. For simplicity, we restrict the following discussion to the
case
where {\it precisely two rays cross at a crossover point}. Throughout the
discussion, we assume that ${\cal S}_0$ is a smooth convex surface embedded at
constant time in Minkowski space. The assumption that precisely two rays cross
at
a point of ${\cal X}$ defines a subset of such convex surfaces which is
invariant under perturbations. This subset includes the case in which ${\cal
S}_0$
is ellipsoidal which is used to construct the
analytical white hole horizons in Sec. \ref{sec:hor} .

Two maps are useful in discussing the generic properties of ${\cal X}$. One is
the
the null geodesic map ${\cal G}_{\hat t}$ from points of ${\cal S}_0$ to points
an affine
time ${\hat t}$ along the generators of $\hat {\cal H}^-$. The crossover set
results from
this map for times ${\hat t}_{\cal X}$ (which vary from generator to
generator). This map
can be extended past the endpoints of $\hat {\cal H}^-$. The other is the
ray map ${\cal R}_{\hat t}$ from ${\cal S}_0$ obtained by applying the
projection
${\cal P}: (\hat t,x,y,z)\rightarrow (x,y,z)$ to the null geodesic map.
${\cal R}_{\hat t}$ is the standard spatial ray-tracing map
of geometric optics based upon the straight Euclidean lines orthogonal to
${\cal
S}_0$. In this projected spatial picture, null rays which cross in the
4-dimensional
sense that they pass through the same space-time point correspond to spatial
rays
from points $p$ and $q$ on ${\cal S}_0$ which intersect at a point an equal
distance
$d$ from $p$ and $q$. Since the rays are normal to ${\cal S}_0$, this implies
the existence
of a sphere centered at ${\cal R}_d(p)={\cal R}_d(q)$ which touches ${\cal
S}_0$ from
the inside at $p$ and $q$. A ray from any point $p$ on ${\cal S}_0$ which does
not
first caustic must cross some other ray, as follows from considering spheres of
increasing radii which touch ${\cal S}_0$ at a given point $p$. The
restriction that triple or higher order crossover points do not occur implies
that such spheres never touch ${\cal S}_0$ at more than two points.
We are especially interested in the generic properties of the double-crossover
set ${\cal X}_D$ consisting of non-caustic points at which precisely two rays
intersect.

{\bf Generic Property 1.}
  {\it A caustic point is not also a crossover point.}
\newline
Let ${\cal R}_{\hat t}$ for $\hat t=c$ map the point $p$ on ${\cal S}_0$ into
the
projection of a caustic point. That implies that ${\cal R}_c(p)$ is the center
of a sphere $S$ of
radius $c$ which osculates ${\cal S}_0$ along its smallest principle curvature
direction at $p$. There are two possibilities. First, suppose there were a
continuous curve of points on ${\cal S}_0$ containing $p$ whose rays all
cross the ray from $p$. Then all points on the curve would be equidistant
from ${\cal R}_c(p)$
and have equal values of their smallest principle radius of curvature. A small
deformation of ${\cal S}_0$ in the neighborhood of $p$ (keeping $p$ and the
radius of curvature at $p$ fixed) would remove this degeneracy.
(The structural instability of this situation also follows directly from the
theory
of generic caustics.) Second, suppose the ray from a point $q$ on ${\cal S}_0$
crosses the ray from $p$, so that the osculating sphere $S$ touches
${\cal S}_0$ at $q$.
Then a small outward deformation of ${\cal S}_0$ in the neighborhood of $q$
would eliminate the crossover. Thus a joint caustic-crossover point can be
eliminated by a perturbation.

{\bf Generic Property 2.} {\it Considered as a subset of Minkowski space, the
double-crossover set ${\cal X}_D$ is a smooth, open, spacelike 2-surface.}
\newline
Let $q$ be a point in ${\cal X}_D$ resulting from the crossing of null rays
from points $p_1$ and $p_2$ on ${\cal S}_0$. Let $\hat t_1$  be the affine time
from $p_1$ to $q$ and consider the portion of null hypersurface ${\cal N}_1$
obtained by the null geodesic map ${\cal G}_{\hat t}$ from ${\cal S}_0$ in a
neighborhood ${\cal O}_1$ of $p_1$ with affine lengths
$(\hat t_1 -\epsilon <\hat t <\hat t_1 + \epsilon)$. Since, for sufficiently
small neighborhoods, there is no caustic in ${\cal N}_1$, this geodesic map is
one-to-one and smooth so that ${\cal N}_1$ is a smooth open portion of null
hypersurface containing $q$. A similar construction based on $p_2$ leads to a
smooth
open portion of null hypersurface  ${\cal N}_2$, which also contains $q$.
Since, at
$q$, the null directions lying in ${\cal N}_1$ and ${\cal N}_2$ are distinct,
we
can choose (by an appropriate boost) a Lorentz frame such that their spatial
components lie in opposite directions, say the $\pm z$ directions. Then, in the
tangent space of $q$, ${\cal N}_1$ and ${\cal N}_2$ intersect in the $(x,y)$
plane.
Since ${\cal N}_1$ and ${\cal N}_2$ are smooth, this guarantees a smooth
spacelike
intersection in some neighborhood of $q$.

{\bf Generic Property 3.}
{\it In the absence of triple or higher order crossovers, the caustic set forms
a
compact boundary to the crossover set (considered as a subset of
$\hat {\cal H}^-$). The tangent space of the crossover set joins continuously
to the tangent space of the null portion of ${\cal H}^-$ at this caustic
boundary.}
\newline
The double crossover set ${\cal X}_D$ lies in a
bounded region so that its boundary must be
compact. Since ${\cal X}_D$ is an open set and the only other endpoints of
${\cal H}^-$ are caustics, the boundary of ${\cal X}_D$ consists of caustics.
Along a curve in ${\cal X}_D$ approaching a caustic point $p$ on the
boundary, the two distinct null directions normal to
${\cal X}_D$ (corresponding to the two crossing rays) have as a common
limit the tangent to the null ray reaching $p$ along
${\cal H}^-$. Hence, the  crossover surface asymptotically
becomes tangent to this null direction at its boundary ${\cal C}$.

These generic properties are violated in the case of a spherical
wavefront, where the crossover set and caustic set both degenerate to a
common point. They are also violated in the prolate spheroidal case,
where the crossover set is a curve of degenerate caustics, which is
bounded at each end by a non-degenerate caustic \cite{toroid}.

These three generic properties of a flat space crossover set are satisfied when
${\cal S}_0$ is a triaxial ellipsoid, as explicitly demonstrated
in Sec. \ref{sec:ellip}. For the
ellipsoidal case, the crossover set is in addition smooth
and connected and consists purely of double crossovers.
More complicated examples would allow higher order crossovers.
However, there is limited complexity to the crossover set arising in Minkowski
space from a generic, smooth, convex surface ${\cal S}_0$.

{\bf Generic Property 4.}
{\it A generic crossover point lies at the intersection of at most 4 rays.}
\newline
Consider a sphere tangent to ${\cal S}_0$ at five or more points,
whose center lies at the projection ${\cal P}$ of the corresponding crossover
point.
Since generically four points determine a unique sphere, an outward
deformation of ${\cal S}_0$ leaving precisely four of these points points
fixed would reduce the crossover to a four-fold intersection.

The conformal construction of the curved space event horizon
(described in Sec. \ref{sec:hor}) preserves the structure
of the the underlying flat space crossover set. The generic properties of
an {\em arbitrary} curved space event horizon are important ingredients of
black hole
physics. Certain of the flat space properties generalize easily to curved
space,
e.g. the spacelike nature of the crossover set. However, for those properties
established using specifically Euclidean constructions,
e.g. Generic Property 4, the generalization is not obvious.

\subsection{The wavefront emanating from an ellipsoid}
\label{sec:ellip}

We now specialize to the case of the ingoing wavefront from an ellipsoid
${\cal S}_0$ described in Cartesian coordinates at time ${\hat t}=0$
by points $x^i=(X,Y,Z)$ satisfying
\begin{equation}
    \frac{1}{2}Q_{ij} X^i X^j :=\frac{X^2}{\Xi^2-a^2}+
    \frac{Y^2}{\Xi^2-b^2}+\frac{Z^2}{\Xi^2}  =1,
\label{eq:define_ellipsoid}
\end{equation}
where we set $\Xi> a> b> 0$ (in the non-degenerate case).
Thus the $x$-axis is the shortest axis, the $z$-axis
is the longest axis and the $y$-axis is intermediate.
The reflection symmetry with respect to the Cartesian axes allows us to
reduce our analysis to the positive octant $x^i\ge 0$. While reflection
symmetry simplifies the analysis, the caustic structure
of the wavefront is generic due to the absence of continuous symmetries, so
that it is preserved under perturbations which break reflection symmetry.

The unit normal to ${\cal S}_0$ is $N_i = Q_{ij} X^j/N$, where
\begin{equation}
          N^2 =\frac{X^2}{(\Xi^2-a^2)^2}+
    \frac{Y^2}{(\Xi^2-b^2)^2}+\frac{Z^2}{\Xi^4}.
\end{equation}
The wavefronts ${\cal S}_{\hat t}$ propagating inward along the rays from
${\cal S}_0$ are given by $x^i=X^i -\hat t N^i$, with $\hat t =const$.
The crossover set
${\cal X}$ formed by these wavefronts must have reflection symmetry and lie in
the plane $x=0$ normal to the short axis of the ellipsoid. Thus the
Minkowski time ${\hat t}_{\cal X}$ (or equivalently the
distance) along a ray from ${\cal S}_0$ to ${\cal X}$ is
\begin{equation}
      {\hat t}_{\cal X} =N(\Xi^2-a^2).
\label{eq:htx}
\end{equation}
The ray with coordinates $(X,Y,Z)$ on ${\cal S}_0$ hits ${\cal X}$ at
\begin{equation}
  x^i =[0,\frac{(a^2-b^2)Y}{(\Xi^2-b^2)},
           \frac{a^2 Z}{\Xi^2}].
\end{equation}
In order to determine the boundary of the crossover set consider rays from the
curve $(X= \epsilon,Y,Z)$. Such a ray will cross with its opposite ray
from  $(X=-\epsilon,Y,Z)$. However, for $\epsilon=0$ the rays are identical
and no such crossover can occur so that these rays form the boundary of the
crossover set.
In the limiting case as $\epsilon \rightarrow 0$ these crossover points
approach the caustic set ${\cal C}$, given by
\begin{eqnarray}
     y &=& \frac{(a^2-b^2)}{\Xi}\sqrt{\frac {\Xi^2-Z^2}{\Xi^2-b^2} } \nonumber
\\
     z &=& \frac{a^2 Z} {\Xi^2}
\label{eq:cell}
\end{eqnarray}
at time
\begin{equation}
   \hat t=\frac{(\Xi^2-a^2)}{\Xi^2}\sqrt{\frac{\Xi^4-b^2 Z^2}{\Xi^2-b^2}}.
\label{eq:tc}
\end{equation}

The caustics are determined by the radii of curvature of ${\cal S}_0$. These
are
computed from the extrinsic curvature tensor of ${\cal S}_0$ which is given by
\begin{equation}
      K_{ij}=(\delta_i^k-N_i N^k)(\delta_j^l-N_j N^l)\partial_k N_l
       =\frac{1}{N}(\delta_i^k-N_i N^k)(\delta_j^l-N_j N^l) Q_{kl}.
\end{equation}
In order to describe the principal curvature directions it is useful to
introduce confocal ellipsoidal coordinates \cite{morsef}, which are determined
by the
three independent solutions to
\begin{equation}
    \frac{x^2}{\xi_i^2-a^2}+
    \frac{y^2}{\xi_i^2-b^2}+\frac{z^2}{\xi_i^2}  =1,
\label{eq:define_ellipsoidal_coords}
\end{equation}
where we set $\xi_i =(\xi,\eta,\lambda)$ with nonsingular coordinate
ranges $\xi>a>\eta >b>\lambda>0$.
Comparison of Eqs. (\ref{eq:define_ellipsoid}) and
(\ref{eq:define_ellipsoidal_coords}) shows that the initial wavefront
${\cal S}_0$ is given by $\xi=\Xi$.

Under the resulting coordinate transformation
\begin{eqnarray}
    x&=&\sqrt{ \frac{(\xi^2-a^2)(a^2-\eta^2)(a^2-\lambda^2)}
              {a^2(a^2-b^2)} } ,\nonumber \\
    y&=&\sqrt{ \frac{(\xi^2-b^2)(\eta^2-b^2)(b^2-\lambda^2)}
              {b^2(a^2-b^2)} } ,\nonumber \\
    z&=& \frac{\xi\eta\lambda}{ab} ,
\end{eqnarray}
the Euclidean metric takes the form
\begin{eqnarray}
    ds^2&=&\frac{(\xi^2-\eta^2)(\xi^2-\lambda^2)}
              {(\xi^2-a^2)(\xi^2-b^2)} d\xi^2 \nonumber \\
         &+&\frac{(\xi^2-\eta^2)(\eta^2-\lambda^2)}
               {(a^2-\eta^2)(\eta^2-b^2)} d\eta^2 \nonumber \\
         &+&\frac{(\eta^2-\lambda^2)(\xi^2-\lambda^2)}
         {(a^2-\lambda^2)(b^2-\lambda^2)}d\lambda^2.
\end{eqnarray}

The principal curvature directions are $\partial_\eta$
and $\partial_\lambda$. The corresponding extrinsic curvature scalars are
\begin{equation}
    K_\eta^\eta =\Xi\sqrt{\frac{(\Xi^2-a^2)(\Xi^2-b^2)}
           {(\Xi^2-\eta^2)^3(\Xi^2-\lambda^2)}}
\label{eq:ketaeta}
\end{equation}
and
\begin{equation}
    K_\lambda^\lambda =\Xi\sqrt{\frac{(\Xi^2-a^2)(\Xi^2-b^2)}
           {(\Xi^2-\eta^2)(\Xi^2-\lambda^2)^3}},
\label{eq:klamlam}
\end{equation}
with $K_\lambda^\lambda \le K_\eta^\eta$.

There are four umbilical points on the ellipsoid where
$K_\lambda^\lambda = K_\eta^\eta$.
In Cartesian coordinates, they are located at the reflection
symmetric set $X=\pm(1/a)\sqrt{(\Xi^2-a^2)(a^2-b^2)}$, $Y=0$, $Z=\pm b\Xi/a$.
According to equation (\ref{eq:htx}), the crossover time for an umbilic ray is
\begin{equation}
     \hat t_{{\cal X},U} =\frac {\sqrt{(\Xi^2-b^2)(\Xi^2-a^2)}}{\Xi}.
     \label{eq:tumbx}
\end{equation}
In confocal coordinates, the umbilical points are at the coordinate
singularity $\lambda=\eta=b$. Since the radii of curvature are the inverse of
the principal curvature scalars, Eq. (\ref{eq:ketaeta}) implies that the focal
length at an umbilic is
\begin{equation}
     \hat t_{{\cal C},U} =\frac {(\Xi^2-b^2)\sqrt{(\Xi^2-b^2)(\Xi^2-a^2)}}
        {(\Xi^2-a^2)\Xi},
   \label{eq:tumbc}
\end{equation}
which is also the caustic time for an umbilic ray. Comparison of Eqs
(\ref{eq:tumbx})
and (\ref{eq:tumbc}) shows that $\hat t_{{\cal C},U} > \hat t_{{\cal X},U}$.
Thus an umbilic ray crosses its opposite umbilic ray before it caustics, except
in the degenerate case $a=b$ of a prolate spheroid. Later we will see that
this has an important consequence for the curved space model of the horizon.
It implies that the mechanism for a pair of eternal black holes found in the
case of a prolate spheroid \cite{ndata} is exceptional and does not occur in
any
other ellipsoidal model.

Also, a calculation of the radius of curvature for rays emanating
from the curve $X=0$ (or equivalently the curve $\eta =a$) shows that they
caustic at the time
\begin{eqnarray}\label{eq:flat_caustic_time}
      {\hat t}_{\cal C}&=& \frac{\Xi^2 -a^2}{\Xi}
    \sqrt{ \frac{\Xi^2-\lambda^2}{\Xi^2-b^2}} \nonumber \\
          &=& \frac{\Xi^2 -a^2}{\Xi^2}
          \sqrt{ \frac{\Xi^4-b^2 Z^2}{\Xi^2-b^2}}.
\end{eqnarray}
Comparison with Eq. (\ref{eq:tc}) shows that this is the limit of the crossover
time for neighboring rays so that this caustic set bounds the crossover set,
i.e.
${\cal C}=\partial{\cal X}$ as expected. Furthermore, from Eq. (\ref{eq:cell}),
the caustic curve ${\cal C}$ is smooth and, in fact, given by the ellipse
\begin{equation}\label{eq:cellips}
     \frac{y^2 (\Xi^2-b^2)}{(a^2-b^2)^2}+\frac{z^2\Xi^2}{a^4}=1.
\end{equation}
in the $x=0$ plane.
Equation (\ref{eq:flat_caustic_time}) shows that the first rays to caustic
are the ones with maximal $Z$ (or equivalently $\lambda =b$),
i.e. the two rays that come from the ``tips'' of the longest extension of
the ellipsoid.

The qualitative features of ${\cal C}$ and ${\cal X}$ are
similar to those for the special case of a wavefront from an {\em oblate}
spheroid, except in that case the caustic curve is a circle of symmetry.
In the {\em prolate} spheroidal case, the
caustic ellipse degenerates into a line on the symmetry axis. The oblate and
prolate spheroidal cases form two pieces of the
boundary of the ellipsoidal state space $\Xi\ge a\ge b\ge 0$ governing
the shape of ${\cal S}_0$. The prolate spheroids correspond to $a=b$ and the
oblate spheroids to $b=0$. These two pieces of the boundary meet at the
spherical case $a=b=0$. The remaining piece of the boundary is formed by the
singular one parameter set of infinitesimally thin ellipsoids
(elliptical pancakes) given by $a=\Xi$, which connects the singular prolate
spheroid (needle) to the singular oblate spheroid (circular pancake) as $b$
ranges from $\Xi$ to $0$. The important feature here is that the interior of
this state space is topologically trivial and should not introduce
discontinuous qualitative behavior such as occurs on the boundary of the
state space in passing from prolate to oblate spheroids.

\section{Conformally ellipsoidal horizons}
\label{sec:hor}

\subsection{Conformal construction of a curved space horizon}

Following Ref. \cite{ndata}, we describe the intrinsic geometry of the horizon
as a stand-alone object possessing all the properties of a non-singular horizon
embedded in a vacuum space-time.  As explained in Ref. \cite{ndata}, the
intrinsic
horizon geometry
provides part of the data to reconstruct the embedding space-time by means of a
double-null characteristic initial value problem
\cite{sachsdn,helmut81a,helmut81b,haywdn}. The white hole horizon ${\cal H}^-$
consists
of the closure of the non-singular portion of a null hypersurface ${\cal N}$
whose surface area has finite asymptotic limit in the past and decreases into
the future.

The construction begins with the 4-dimensional description of ${\cal N}$ as a
null
hypersurface embedded in a vacuum space-time with metric $g_{ab}$, covariant
derivative $\nabla_a$ and an affine tangent $n^a$ to the null generators of
${\cal
N}$. We make no assumptions about the behavior of $n^a$ off ${\cal N}$. On
${\cal
N}$, it satisfies the geodesic equation $n^b\nabla_b n^a=0$ and the
hypersurface
orthogonality condition  $n^{[a}\nabla^b n^{c]}=0$. The choice of $n^a$ has
freedom
$n^a\rightarrow \alpha n^a$ and determines an affine parameter $u$ satisfying
$n^a
\partial_a =\partial_u$ with the freedom $u\rightarrow \alpha^{-1} u+\beta$
(with
$\alpha$ and $\beta$ constant along the generators).

{}From this embedding picture, we induce an intrinsic 3-dimensional
description of ${\cal N}$
by projecting 4-dimensional tensor fields into ${\cal N}$. Although this cannot
be done by the standard ``3+1'' decomposition used for a spacelike
hypersurface, we base an analogous approach upon the projection operator
\begin{equation}
       P_a^b = \delta_a^b + n_a l^b,
\end{equation}
where $l_a = -\nabla_a u$, with $u$ any smooth extension of the affine
parameter
field on ${\cal N}$. Then $n_a=g_{ab}n^b$ projects to 0; the projected metric
$\gamma_{ab}= P_a^c P_b^d g_{cd}$ is the intrinsic metric of ${\cal N}$ with
the
degeneracy $\gamma_{ab} n^b =0$; and, restricted to the surfaces $u=const$, the
projected contravariant metric $\gamma^{ab}= P^a_c P^b_d g^{cd}$ is the inverse
of
the pullback of $\gamma_{ab}$. We set $\gamma_{ab}=R^2 h_{ab}$ and
$\gamma^{ab}=R^{-2} h^{ab}$ where $h^{ab} {\cal L}_n h_{ab} =0$. (This is
achieved
by factoring out the square root of the determinant of the restriction of
$\gamma_{ab}$ to the
surfaces of the affine foliation such that $h_{ab}$ is unimodular.)
The Lie derivative of $h_{ab}$ along the
generators of ${\cal N}$ defines the shear tensor
$\Sigma_{ab}={\cal L}_n h_{ab}$.

Let $\perp T_a^b$ denote the projection
of a tensor $T_a^b$ to the tangent space of ${\cal N}$.
The projected curvature components
$\Phi_{ab}=\perp n^c n^d R_{cabd}=\perp n^c (\nabla_c \nabla_a-\nabla_a
\nabla_c) n_b$, can be re-expressed in a form intrinsic to ${\cal N}$ as
\begin{equation}
  \Phi_{ab} = \frac{1}{2}{\cal L}_n (R^2 h_{ab})
        +h_{ab}R {\cal L}_n^2 R -  \frac{1}{4}R^2
   h^{cd}\Sigma_{ac}\Sigma_{bd}.
  \label{eq:phiab}
\end{equation}

The vacuum Einstein equations require that the trace of $\Phi_{ab}$
vanish, leading to the focusing equation
\begin{equation}
   {\cal L}_n^2 R = -\frac{1}{4} R \Sigma^2,
\label{eq:focus}
\end{equation}
where $\Sigma^2 =(1/2) h^{ab}h^{cd}\Sigma_{ac}\Sigma_{bd}$. Also,
the trace free part of Eq. (\ref{eq:phiab}) yields
\begin{equation}
       \Psi_{ab} = \frac{1}{2}{\cal L}_n (R^2 \Sigma_{ab})
                 - \frac{1}{2}R^2 \Sigma^2 h_{ab}.
\label{eq:psiab}
\end{equation}
where $\Psi_{ab}=\perp n^c n^d C_{cabd}$
are projected components of the Weyl curvature.

The normalized eigenvectors $p^a$ and $q^a$ of the shear tensor,
satisfying $\Sigma_{ab}(p^b+iq^b)=\Sigma h_{ab}(p^b-iq^b)$, provide a
basis in which
$h_{ab}=p_ap_b+q_aq_b$ and $\Sigma_{ab}=\Sigma (p_a p_b-q_aq_b)$, where
$p_a+iq_a=h_{ab}(p^b+iq^b)$. In this basis, the Weyl curvature components
$\Psi_{ab}$ can be represented by the complex scalar field
\begin{equation}
   \Psi = R^{-2}(p^a +iq^a)(p^b +iq^b)\Psi_{ab}
    =R^{-2}{\cal L}_n (R^2 \Sigma)
     +i\Sigma(q^a{\cal L}_n p_a -p^a{\cal L}_n q_a).
\label{eq:Psi}
\end{equation}

The conditions for a non-singular horizon ${\cal H}^-$ require that
$\Psi$, $R$ and $h_{ab}$ be smooth fields and that $R$ has a finite
limit $R_\infty$ as $u\rightarrow -\infty$.
Reference \cite{ndata} presents a conformal method for solving the
focusing equation consistent
with these horizon regularity conditions.
This solution restricts the outgoing
radiation crossing ${\cal N}$ by requiring that its intrinsic metric be
conformal to that of a null hypersurface embedded in a flat Minkowski
space-time.

We denote the corresponding
flat space fields on ${\cal N}$ as ${\hat \gamma}_{ab}$, $\hat R$,
${\hat h}_{ab}$, $\hat n^a$, $\hat u$, etc. For convenience, we write
$F'= {\cal L}_{\hat n} F$ for tensor fields $F$. Since $\hat \Psi
=0$, Eq. (\ref{eq:psiab}) implies $(\hat p^a+i\hat q^a)(\hat p_a-i\hat
q_a)'=0$ and ${\hat R}^2 \hat \Sigma =\sigma$, where $\sigma '=0$. The
conditions on the eigenvectors are
$\hat p_a '=(\hat \Sigma /2)\hat p_a$ and
$\hat q_a '=-(\hat \Sigma /2)\hat q_a$.

It is convenient to adjust the affine freedom in $\hat u$ so that
$\hat R\rightarrow -\hat u$ as $\hat u \rightarrow -\infty$
and so that the two caustics encountered along each ray (where $\hat R =0$)
are placed symmetrically. Then the flat space version of
the focusing equation (\ref{eq:focus}) integrates to give
\begin{equation}
 \hat R^2=(\hat u+\frac{1}{2}\sigma)(\hat u-\frac{1}{2}\sigma).
\label{eq:rhat}
\end{equation}

In the flat embedding, $\sigma$ is the distance between the two caustics
generically encountered along each null ray.
(We use the convention $\sigma \ge 0$, so that the
caustic corresponding to the $\hat q$ principle direction is reached
first, moving along a ray in the direction of increasing $\hat u$.)
The eigenvectors have the form
\begin{equation}\label{eq:pa_solution}
\hat p_a   = \bigg( \frac {\hat u -\sigma /2}
                   {\hat u +\sigma /2} \bigg)^{1/2} P_a
\end{equation}
\begin{equation}\label{eq:qa_solution}
\hat q_a   = \bigg( \frac {\hat u +\sigma /2}
                   {\hat u -\sigma /2} \bigg)^{1/2} Q_a
\end{equation}
where $(P_a+iQ_a)'=0$.

In applying this construction, we generate the flat space null hypersurface
by the ingoing light rays from a convex
surface ${\cal S}_0$ embedded at time $\hat t =0$ in Minkowski space-time.
We construct the curved space horizon with the same conformal structure, i.e.
$h_{ab}={\hat h}_{ab}$, by setting  $\gamma_{ab}=\Omega^2\hat \gamma_{ab}$
with $R=\Omega \hat R$. The associated
affine structures are related by $n^a=\Lambda \hat n^a$ so that
$\partial_u = \Lambda \partial_{\hat u}$ and $\Sigma =\Lambda \hat
\Sigma$.

The curved space focusing equation (\ref{eq:focus}) then
reduces to
\begin{equation}
     \Lambda' (\Omega'  \hat R  +\Omega  \hat R')
    +\Lambda  (\Omega'' \hat R +2\Omega' \hat R') = 0,
\label{eq:foci}
\end{equation}
with the Weyl scalar, defined in Eq. (\ref{eq:Psi}), given by
\begin{equation}
       \Psi = \sigma \frac{ \Lambda (\Omega^2\Lambda)'}{\Omega^2 \hat R^2}.
\label{eq:psi}
\end{equation}

The ansatz
\begin{equation}
    \Omega=-R_{\infty}\big( \hat u
          +\frac{\sigma^2}{12(\rho-\hat u )}\big)^{-1}
\label{eq:ansatz}
\end{equation}
satisfies the horizon regularity requirements provided $\rho$ is a
suitably chosen function on ${\cal S}_0$, as discussed below.
The simplicity of this ansatz suggests that it does not introduce an
extraordinary amount of radiation crossing the horizon.

Integration of Eq. (\ref{eq:foci}), with the affine scale fixed by
the condition $u'=1/\Lambda \rightarrow 1$
as $\hat u \rightarrow -\infty $, gives
\begin{equation}
    u'= 1/\Lambda =\frac{9}{(12 \hat u (\hat u-\rho) -
      \sigma^2)^2} \frac{( 5 \rho + \mu-2 \hat u)^{2 \,
          (2\rho/\mu +1) \,}}{( 5 \rho- \mu-2 \hat u)^{2 \,
          (2\rho/\mu -1) \,} } \, ,
\label{eq:lampr}
\end{equation}
where
\begin{equation}\label{eq:def_mu}
\mu = \sqrt{13\rho^2 -\sigma^2}.
\end{equation}
On ${\cal S}_0$,
we fix the remaining affine freedom in $u$ by setting $u=\hat u =u_0$.  The
$S_{\hat t}$ foliation of ${\cal N}$ defined by the parameterization
$\hat u =u_0 +\hat t$ is the standard
geometric optics family of wavefronts obtained from ${\cal S}_0$ by Huyghen's
construction. We adopt the $S_t$ foliation of the horizon defined by $u=u_0+t$
to define the evolution of the white (or black) holes. While this definition
is somewhat arbitrary, it is based upon a foliation intrinsically related to
the horizon structure so that it does not introduce an artificially distorted
picture of the bifurcation process.

Equation (\ref{eq:lampr}) now determines the deviation of the
new slicing which is adapted to the curved spacetime from the original
slicing adapted to the embedding of the null surface in Minkowski spacetime.
This equation thus controls the change of topology of spatial slices of the
event horizon.

Some qualitative understanding of the resulting effects can be gained from a
simple
analysis of Eq. (\ref{eq:lampr}).
We present this analysis in three steps. First, we analyze the
regularity of the curved horizon, which gives an algebraic
restriction on the function $\rho$.
Second, we consider the significance of umbilical rays for which $\sigma=0$.
Third,  we discuss general features
of the function $\Lambda$, which determines the ``dynamics'' of the white hole
and provides a simple picture of the relative rate between the $t$ and $\hat t$
foliations, as determined by Eq. (\ref{eq:lampr}).
This helps to understand the main features
of the topology change phenomena, which will be laid out in Sec.
\ref{sec:results}.

As seen from Eq. (\ref{eq:psi}), $\Psi$ is potentially singular at caustic
points where $\hat R =0$. This singularity is of no concern for a ray whose
caustics lie outside ${\cal H}^-$. However, a non-singular horizon requires
that $\Psi$ be non-singular at those caustics which lie on the endpoints of
${\cal H}^-$.
The first caustic is reached at $\hat u =-\sigma/2$ along each ray. Inserting
our ansatz (\ref{eq:ansatz}) and Eq. (\ref{eq:lampr}) for $\Lambda$, the Weyl
scalar at such points becomes
\begin{equation}
   \Psi=\frac{8\sigma^4}{81(\sigma+2\rho)^6 (\sigma+3\rho) }
       \bigg (\frac{5\rho-\mu+\sigma}{5 \rho+\mu+\sigma }\bigg )^{8\rho/\mu}.
     \label{eq:psix}
\end{equation}
In order to have $\mu$ non-negative and $\Omega$ strictly positive on
${\cal H}^-$ one has to require
\begin{equation}\label{eq:rho_inequality}
\rho \ge \sigma/\sqrt{13}.
\end{equation}
Together with Eq. (\ref{eq:psix}) this condition implies that $\Psi$
is smooth except possibly on rays with $\sigma=0$ in the case when $\rho$ also
vanishes
on these rays.

Rays with $\sigma =0$ arise at umbilical points of ${\cal S}_0$ where its two
curvature eigenvalues are equal.
Umbilic points are to be expected on a generic convex surface. Indeed,
Carath{\'e}odory conjectured that every smooth, compact, convex surface
(embedded in a Euclidean 3-space) must have
at least two umbilics \cite{spivak}, which is the number of umbilics for a
prolate or oblate
spheroid. A generic ellipsoid has four umbilics.

The shear $\hat \Sigma$ vanishes along such
rays and thus $\Sigma =0$ since vanishing shear is a conformally invariant
property. So $R=R_{\infty}$ along an umbilic ray,
as a consequence of Eq. (\ref{eq:focus}) and the bounded surface
area of the horizon. This is consistent with our ansatz which reduces to
$\Omega = R_{\infty}/\hat R$ along an umbilic ray.
The completeness of ${\cal H}^-$ as a white hole horizon requires that the rays
extend interminably or to a caustic or crossover point,
whichever occurs first. Along a non-umbilic ray, focusing  ensures that a
caustic is reached at a finite affine time so that the range of $u$ has an
upper bound
on the horizon. However, along an umbilic ray there is no focusing and the
white hole may extend to infinite $u$. This is the mechanism which leads to
eternal black holes in a $t$-foliation of ${\cal H}^+$ for the axisymmetric
head-on collision corresponding to the choice of a prolate spheroid for
${\cal S}_0$ \cite{ndata}. In the prolate spheroidal model of a black hole
collision,
regularity of $\Psi$ was ensured by requiring that $\rho \ge \sigma_M
/\sqrt{13}$, where $\sigma_M$ is the maximum value attained by $\sigma$ on all
the rays generating the horizon.

This places
no extra regularity conditions on the conformal model when the
umbilic rays attain their caustic beyond the endpoints of ${\cal H}^-$.
As already discussed in Sec.~\ref{sec:ellip}, this is the
case for the umbilic rays on an ellipsoidal wavefront.
Only in the degenerate limit of the prolate spheroidal case does
the umbilic caustic approach an endpoint  of the horizon.
In the generic case, the restriction $\rho \ge \sigma_M /\sqrt{13}$ can be
relaxed to allow models with
 $\rho =k\sigma /\sqrt{13}$, with constant $k\ge 1$, which would otherwise
be singular if an umbilic caustic were to lie on ${\cal H}^-$.

\begin{figure}[htb]
\begin{center}
 \epsfxsize=3.375in\leavevmode\epsfbox{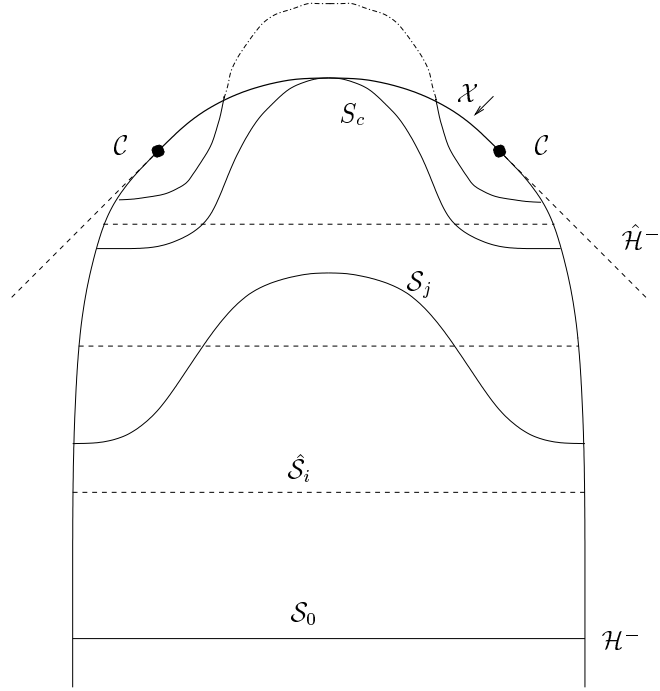}
\caption{Ellipsoidal ($\hat{\cal H}^{-}$) and conformally ellipsoidal
(${\cal H}^{-}$) null hypersurfaces: Suppressing one dimension
allows the foliations to be depicted as lines from pole to pole. The
Minkowski foliation indicated by $\hat S_i$ is drawn horizontally. The curved
space
foliation is indicated by $S_j$. In the top-most slice of this foliation the
white hole has two disjoint components (the part of the slice which is
off the horizon is shown as $- \cdot - \cdot -$).}
\label{fig:slicings}
\end{center}
\end{figure}

In order to elucidate the features of the model we now discuss
how Eq. (\ref{eq:lampr}) determines the evolution of the
white hole, defined by the $S_t$ foliation, relative
to the Minkowski foliation $S_{\hat t}$.
A schematic diagram of the situation is given in Fig. \ref{fig:slicings}.
The affine
parameters $t$ and $\hat t$ are defined so that $t=\hat t = 0$ on
the surface ${\cal S}_0$. These affine parameters are related
to $u$ and $\hat u$ by a shift in origin, $t = u - u_0$ and
$\hat t = \hat u - u_0$. Thus $dt/d\hat t=du/d\hat u=u'$ so that the
relative rate between the $S_t$ and $S_{\hat t}$ foliations is determined by
Eq. (\ref{eq:lampr}).
We first note that Eq. (\ref{eq:lampr}) implies $du/d\hat u > 0$,
with $du/d\hat u \rightarrow 1$ as $\hat u\rightarrow -\infty$. Thus
$u$ is a monotonically increasing function of $\hat u$ so that the
transformation
from $\hat u$ to $u$ is non-singular along each ray, independent of the
values of $\sigma$ and $\rho$.

Furthermore, for our particular ansatz, it is easy to show that
$1/\Lambda =du/d\hat u \ge 1$.
Thus, to the future of ${\cal S}_0$, $\hat t \leq t$
and the Minkowski slicing always ``lags behind''. The ray dependence of this
effect, which depends on the values of $\rho$ and $\sigma$, creates
a tidal bulge, as shown in Fig. \ref{fig:slicings}. This tidal bulge
leads to a change of topology of the white hole just after the
bulge reaches the crossover set ${\cal X}$, corresponding to the critical slice
$S_c$ in Fig. \ref{fig:slicings}.
As will be demonstrated in Sec. \ref{sec:results}, the bulge typically results
in
punching a hole in the white hole, producing a toroidal white hole with a
sharp edge on the inner rim defined by the crossover points.

In order to discuss how this effect varies from ray to ray and how it varies
in time, let us assume that $|u_0| \gg \sigma$ so that ${\cal S}_0$ is located
at
an early in the evolution when the white hole is quasi-stationary. At this
early time
$\Lambda\approx 1 - 12\rho/|\hat u| + (100 \rho^2 -2\sigma^2 /3)/{\hat u}^2
                                                            + O(\hat u^{-3})$
is fairly uniform from ray to ray for models in which $\rho=const$.
But at the caustic, where $\hat u =-\sigma/2)$, Eq. (\ref{eq:lampr}) gives
\begin{equation}
\Lambda=\frac{\sigma^2}{9(\sigma + 2 \rho)^2}\Bigg(\frac{\sigma +5 \rho - \mu}
    {  \sigma+5 \rho + \mu} \Bigg )^{4\rho/\mu} \, ,
\end{equation}
which displays a strong minimum near an umbilical ray if $\rho$ is nonzero.
For $\rho=const.$ this is the dominant effect at late times in the
``quasi-prolate'' case where the umbilic caustic lies close to the horizon.

As a result, in that case the tidal bulge lies in the region between the
umbilics at the critical late stage where it contributes to topology change.
In particular, it then takes an infinite time for the umbilic
rays to reach the caustic in these cases. This is the mechanism which produces
eternal black holes when the crossover set meets the caustic set on an
umbilical
ray. Within the ellipsoidal models this only happens for the prolate spheroids.

If, on the other hand, we choose $\rho$ of the form
$\rho = k\sigma/\sqrt{13}$, no such anomalous behavior occurs near the
umbilics.
In this case, $\Lambda=1$ along a umbilical ray. In regions of large $\sigma$,
this results in an upward bulge of the $t$-foliation,
as shown in Fig. \ref{fig:slicings}.
Some of these models lead to convenient simplification of the analytic
dependence governing
the curved space affine parameter in Eq. (\ref{eq:lampr}). In particular, the
case $\rho=\sigma /3$ leads to the simple result
\begin{equation}
        u'=\frac{(6\hat u -7\sigma)^4}
                {9(12\hat u^2-4\hat u \sigma -\sigma^2)^2}
\label{eq:slamp}
\end{equation}
with integral
\begin{equation}
   u=\hat u +4\sigma \log \bigg (\frac
                    {\sigma-2u_0}{\sigma-2\hat u} \bigg )
      +\frac{16\sigma^2(\hat u-u_0)}{9(\sigma-2\hat u)(\sigma-2 u_0)}
      +\frac{256\sigma^2(\hat u-u_0)}{(\sigma+6\hat u)(\sigma+6 u_0)}
\label{eq:su}
\end{equation}
(where the integration constant has been adjusted so that
$u= \hat u =u_0$ on ${\cal S}_0$).
We have $u=\hat u$
along a ray on which $\sigma =0$ but otherwise it is easily seen
that the additional terms on the
right hand side of Eq. (\ref{eq:su}) imply $du/d\hat u >1$.

More generally, the affine parameterization
is determined by integrating Eq. (\ref{eq:lampr}) numerically
(see Sec. \ref{sec:tech} for further notes).


\subsection{The conformally ellipsoidal case}
\label{sec:axi}

Following the formalism of Sec. \ref{sec:ellip}, we describe ${\cal S}_0$ in
ellipsoidal coordinates by the surface $\xi=\Xi$, with $(\eta,\lambda)$
coordinatizing the lines of principle curvature corresponding to the
$(\hat q^a,\hat p^a)$ directions, respectively. The corresponding principal
radii of curvature,
$r_\eta=1/K_\eta^\eta$ and $r_\lambda=1/K_\lambda^\lambda$, are determined from
Eqs. (\ref{eq:ketaeta}) and (\ref{eq:klamlam}), and satisfy $r_\lambda\ge
r_\eta$. This allows us to calculate
\begin{equation}
       \sigma=r_\lambda- r_\eta =
      \frac{\eta^2-\lambda^2}{\Xi}\sqrt{\frac
           {(\Xi^2-\eta^2)(\Xi^2-\lambda^2)}{(\Xi^2-a^2)(\Xi^2-b^2)}}
\label{eq:sigma}
\end{equation}
and
\begin{equation}
       u_0=-(r_\lambda+ r_\eta)/2 =
      -\frac{2 \Xi^2-\eta^2-\lambda^2}{2 \Xi}\sqrt{\frac
           {(\Xi^2-\eta^2)(\Xi^2-\lambda^2)}{(\Xi^2-a^2)(\Xi^2-b^2)}}.
\label{eq:u0}
\end{equation}
Also, $\sigma$ attains its maximum
\begin{equation}
      \sigma_M =
      \frac{2 \sqrt{3}\,\Xi^3}{9\sqrt{(\Xi^2-a^2)(\Xi^2-b^2)}}
\label{eq:sigmaM}
\end{equation}
at $(\eta=\Xi\sqrt{2/3},\lambda=0)$.

Since $r_\lambda\geq r_\eta$, the
$\partial/\partial \lambda$-direction corresponds to the
${\hat p}_a$ principal direction in Eq. (\ref{eq:pa_solution}), while the
$\partial/\partial \eta$-direction corresponds to the
${\hat q}_a$ principal direction in Eq. (\ref{eq:qa_solution}). Inserting
the ellipsoidal results into Eqs. (\ref{eq:pa_solution}) and
(\ref{eq:qa_solution}) yields
\begin{eqnarray}\label{eq:PQ}
P_a &=& \sqrt{\frac{(\eta^2-\lambda^2) (\Xi^2 - \eta^2 )^3}
         {(a^2-\lambda^2)(b^2-\lambda^2)(\Xi^2-\lambda^2)}}
\,\,(d\lambda)_a, \nonumber \\
Q_a &=&  \sqrt{\frac{(\eta^2-\lambda^2) (\Xi^2 - \lambda^2 )^3}
               {(a^2-\eta^2)(\eta^2-b^2)(\Xi^2-\eta^2)}}\,\,(d\eta)_a.
\end{eqnarray}
The time dependent metric for the curved slicing is then given by
\begin{equation}\label{eq:the_intrinsic_solution}
\gamma_{ab} = \Omega^2 {\hat R}^2 \left
               ({\hat p}_a{\hat p}_b + {\hat q}_a{\hat q}_b
                                  \right),
\end{equation}
where the functions $\Omega$, $\hat R$, ${\hat p}_a$ and ${\hat q}_a$ are
computed via Eqs. (\ref{eq:rhat}) - (\ref{eq:qa_solution}), (\ref{eq:ansatz})
and (\ref{eq:PQ}). Here the dependence on the arguments
$\eta$, $\lambda$ and $\hat u$ is known explicitly but, in order to
determine the dependence on the curved slicing,
$\hat u(u)$ has to be computed numerically in general.
This procedure only determines the intrinsic geometry of the null surface.
Its extrinsic curvature in the physical spacetime is determined independently
by an additional constraint equation, which we do not consider here.

\subsection{Numerical Construction and Visualization}
\label{sec:tech}

A particular model for the horizon is determined by a specification of
the ellipsoid to serve as the surface ${\cal S}_{0}$ and a
specification of the conformal factor
$\Omega$. Fixing the ellipsoid corresponds to choosing the parameters
$\Xi$, $a$ and $b$ (where of course, one could eliminate one of these to
fix the scale). Fixing $\Omega$ amounts to a choice of
the functions $R_\infty$ and $\rho$ in our ansatz for the conformal factor in
Eq. (\ref{eq:ansatz}). Here we only consider the case $R_\infty =const$,
independent of ray, which would be satisfied by an asymptotic Schwarzschild
horizon of mass $R_\infty/2$. (In order to model a Kerr horizon, $R_\infty$
would be chosen as the conformal factor relating the intrinsic geometry
of a Kerr white hole to the unit sphere.) We consider several choices
of $\rho$, subject to the inequality (\ref{eq:rho_inequality}).

The crossover and caustic times in terms of the ``curved''
affine parameter $t(\hat t)$ are obtained via the integration
\begin{equation}\label{eq:integrate_for_u}
t(\hat t) = \int_{0}^{\hat t}\frac{d\hat t}{\Lambda(\hat t +u_0)}
\end{equation}
along each ray, where $\Lambda$ is given by Eq. (\ref{eq:lampr}).
As has been discussed in the last section, for certain choices of the
function $\rho$ in the ansatz for the conformal
factor in Eq. (\ref{eq:ansatz}), such as $\rho=\sigma/3$, this integration
can be carried out exactly. For generic choices, the integrations have been
performed using the routines {\em dqag} and {\em qag} from the freely available
package QUADPACK \cite{QUADPACK}.

In addition one also wants to study the foliation $S_t$ of the horizon,
e.g. to animate the surface and show its change of topology.
This task requires the integration of the ODE
\begin{equation}
\frac{d\hat t}{dt} = \Lambda(\hat u) =\Lambda(\hat t + u_0)
\label{eq:ODE}
\end{equation}
to determine $\hat t(t)$
along each ray, which is the ``inverse'' of the integration in
Eq. (\ref{eq:integrate_for_u}).
Even for cases when a special choice of $\rho$ leads to an explicit function
$t(\hat t)$, inversion to $\hat t(t)$ can still not be carried out
analytically but one can use a simple root finding algorithm. In the general
case, the ODEs have to be solved numerically. This has been done using
Runge-Kutta methods implemented in the freely available package
rksuite90 \cite{rksuite90_1,rksuite90_2}, which provides an efficient solution
of a system of ODEs by using adaptive step-size control and data
structures that allow straightforward vectorization.

Visualization of a slice of the horizon is not straightforward
since an isometric embedding in $R^3$ does not exist in general.
However, for present purposes, our primary aim is only to visualize the
topology of these surfaces and the sharpness that occurs
at the crossover points. Since locally the metrics of any two surfaces
are conformally equivalent, this amounts to suppressing the
information contained in the conformal factor. (In a more detailed picture,
the conformal factor could be color-coded onto the surface to
visualize the actual metric geometry to some extent.)

Here we use a very simple approach based on a natural visualization of the
foliation $S_{\hat t}$ of the original null hypersurface $\hat {\cal H}^-$ in
Minkowski space (introduced in Sec. \ref{subsec:flat:general properties} in
terms of the boundary of the future of ${\cal S}_0$).  A spatial reconstruction
of the wave front at a given time $\hat t$ is easily obtained by projecting
$\hat {\cal H}^-$ into the $\hat t=0$ Euclidean 3-space, using the  Minkowski
space projection map ${\cal P}$ (also introduced in Sec.
\ref{subsec:flat:general properties}). Because $\hat {\cal H}^-$ is achronal
this map is one-to-one and preserves the smoothness of curves and surfaces.
Applied to a sequence of slices $S_{\hat t}$, the map ${\cal P}$ generates a
movie of the wave front's motion in space  which faithfully reproduces the
topology of the foliation and does not introduce any extraneous sharp edges.

Our concern is the corresponding projection map of the white hole foliation
$S_t$ into the $\hat t=0$ Euclidean 3-space. The affine parameter $t$ is a
smooth monotonically increasing function of $\hat t$ along each ray up to the
caustic and crossover points.  This property extends to the crossover points
because of the reflection symmetry of the initially ellipsoidal null
hypersurface. The reflection symmetry guarantees consistency of the values of
$u$ when
non-neighboring rays cross so that the crossover set of the curved space
horizon can be
consistently identified pointwise with that of the flat-space null
hypersurface.
Accordingly, $t$ is single valued function on the crossover
set and the transition from $S_{\hat t}$ to $S_t$ does not introduce any
extraneous sharp edges. The projection ${\cal P}$ applied to $S_t$ again
provides a faithful visualization of topology and smoothness, in this case of
the white hole foliation.

\section{The bifurcation of an ellipsoidal horizon}
\label{sec:results}

The preceding analysis describes the dependence of the null geometry of ${\cal
H}^-$ on the affine parameter along a given ray. In order to model a
fissioning white hole we now examine the global dependence of the geometry on
the coordinates parameterizing the rays of ${\cal H}^-$ for an ellipsoidal
choice of ${\cal S}_0$.

The application of our approach and the interpretation of the results regarding
topology change of the event horizon require  considerable care. Since the
crossover surface is spacelike, although asymptotically null at its boundary,
we
could in particular choose a slicing which aligns with the crossovers almost
everywhere on a single slice. The topology change in such a slicing would then
be restricted to a region near the caustic boundary. This choice of slicing
would however be non-generic and small perturbations in the slicing would
effectively advance some portions off the horizon, while retarding others.
Under
such a perturbation, the slicing of the event horizon would then look
arbitrarily complicated and, in particular, there would be no intrinsically
defined number of components. See Ref. \cite{siino2} for a similar discussion.
In contrast,
our interest is in generating generic slices, which are expected to resemble
situations one might encounter in a numerical construction of the
spacetime. In a Cauchy evolution one chooses a slicing adapted to the physics
of the
problem which, in the case of a black hole merger, would contain exactly
two components of the event horizon
at early times. Similarly, when using the solutions produced
here as initial data for the numerical construction of the spacetime via the
characteristic initial value problem \cite{highp,wobb}, one also desires a
useful slicing which is
naturally adapted to the physics. The condition we adopt here to
model a binary black hole merger (as opposed to a single, dynamically evolving
black
hole) is that the individual components have
lifetimes lasting much longer than the phase of topology change. This of course
is
not a precise definition, but rather a practical guideline.

For the simplest case when ${\cal S}_0$ is a unit sphere, $\sigma =0$ on all
rays, $\hat h_{ab}$ is the unit sphere metric, $\hat R =-\hat u$ and the ansatz
Eq. (\ref{eq:ansatz}) reduces to $R=R_{\infty}$ so that the horizon is
stationary. Note that the limiting differential equation for $\Lambda$,
obtained from Eq. (\ref{eq:foci})as $\sigma\rightarrow 0$, {\em does} depend on
the
value of $\rho$ in the ansatz (\ref{eq:ansatz}), so that different results
along
an umbilic ray are allowed. These choices will however only affect the
parameterization, not the geometry, since always $R=R_{\infty}$ in this case.
The simple
spherical limit $\Lambda=1$ and thus $t = \hat t$ is guaranteed by choosing the
function $\rho$ to vanish in the limit.

The next simplest case, which is still degenerate and thus also corresponds to
a
singular limit of our model, is when ${\cal S}_0$ is a spheroid. In Ref.
\cite{ndata} we
explicitly worked out the details of the conformal model for this case,
ensuring
smoothness by restricting the ansatz Eq.(\ref{eq:ansatz}) by $\rho =
\sigma_M/\sqrt{13}$, where $\sigma_M$ is the maximum value of $\sigma$ attained
on ${\cal S}_0$. Here we also consider other choices, such as the form
$\rho=const\times \sigma$, in which $\rho$ is not necessarily constant from ray
to ray. As explained in the previous section, such models develop a singularity
in the limit of a prolate spheroid but yield a smooth horizon otherwise. All of
these choices in particular yield the simple spherical limit mentioned
above.

\begin{figure}[htb]
\begin{center}
 \epsfxsize=3.375in\leavevmode\epsfbox{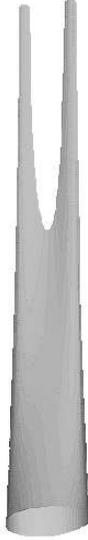}
\caption{The ``pair of pants'' upside down -- corresponding to a fissioning
white hole. The image shows a $x=0$ slice of the null surface
generated from the parameter set BBH (see table \ref{table:parameters})
and was produced with the method described in Sec. \ref{sec:tech}. Spatial cuts
of
this null surface are shown in Figs. \ref{fig:BBH20} -- \ref{fig:BBH46}.}
\label{fig:pop}
\end{center}
\end{figure}

The axisymmetric horizons fall into two qualitatively
different classes, corresponding to the prolate and oblate spheroids.
In axisymmetry we may suppress the angle about the axis of symmetry and display
time vertically to yield a three-dimensional spacetime picture.
In the {\em prolate} case we thus get the well known ``pair-of-pants'' picture
shown in Fig. \ref{fig:pop}, where the seam of the pants is the degenerate
caustic-crossover line and the legs are infinitely stretched.
(Fig. \ref{fig:pop} also encompasses the oblate spheroidal and
generic ellipsoidal cases and was generated from the parameter set BBH,
defined in Table I.)
A similar picture was featured on the cover of Science magazine \cite{science},
where it was described in detail, and also more recently within the
current approach in \cite{ndata}.
This behavior cannot be generic because generically the crossover
surface is 2-dimensional. However, it shows that the choice of
$\rho = const$ yields the same qualitative features found in
the conventional approach of locating the horizon in a given spacetime.

In the oblate spheroidal case the pair of pants picture is still
qualitatively valid if the suppressed dimension of axisymmetry is identified so
that the seam corresponds to a surface of revolution \cite{toroid}.
Whereas in the prolate spheroidal case the seam on the pair-of-pants lies on
the
axis of symmetry, now the crossover surface forms in the equatorial plane and
is indeed 2-dimensional. The spacelike slices ${\cal S}_t$
intersecting it exhibit a toroidal
horizon with sharp inner boundary. Toroidal horizons at the early time
of black hole formation were first found in numerical simulations of the
collapse of an axisymmetric rotating cluster \cite{torus}. Choosing an oblate
spheroidal model with $\rho = const$ produces a similar toroidal horizon within
our ansatz \cite{ndata}. The torus
exhibits a sharp inner ring where it hits the crossover surface. Note that
the whole crossover surface itself is a smooth surface in spacetime. The white
hole vanishes at the moment the torus shrinks to a circle. Due to the
axisymmetry, the torus shrinks with the same speed at all azimuthal angles. In
the
generic case, it would be expected to shrink faster in some places than
others, so that the torus would break apart before the white hole
vanishes. Thus more than one component will appear in the early stages of the
corresponding (time reversed) black hole if we were to perturb the
slicing or the geometry away from
axisymmetry. This does {\em not necessarily} mean that an observer would
interpret
this as
the collision of multiple black holes, because the lifetimes of the individual
components might be shorter than the time necessary to form the
final single black hole.

\begin{figure}[htb]
\begin{center}
 \epsfxsize=3.375in\leavevmode\epsfbox{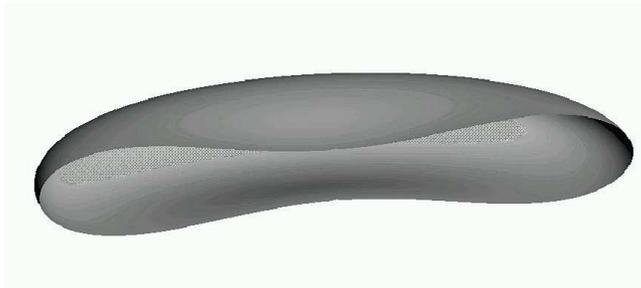}
\caption{Spatial cut of the event horizon for the model QS
at affine time $t= 1.78$.
At this early stage, the wavefront corresponds qualitatively to the
oblate spheroidal case.
The surface is clearly bulged inwards, so the rays at the poles of the event
horizon cross first, but the deviation from axisymmetry is not yet apparent.}
\label{fig:SP31}
\end{center}
\end{figure}

\begin{figure}[htb]
\begin{center}
 \epsfxsize=3.375in\leavevmode\epsfbox{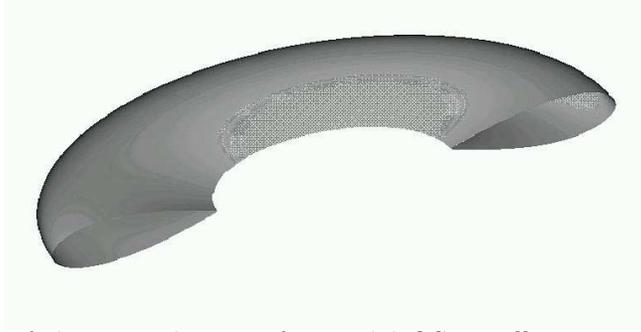}
\caption{Spatial cut of the event horizon for model QS at affine
time $t= 1.90$.
After the rays at the poles of the event
horizon have crossed, the event horizon becomes a torus,
losing points at a sharp inner edge which is a cut of the crossover
surface.}
\label{fig:SP34}
\end{center}
\end{figure}

\begin{figure}[htb]
\begin{center}
 \epsfxsize=3.375in\leavevmode\epsfbox{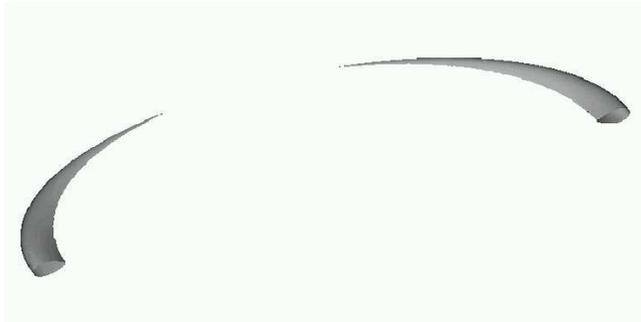}
\caption{Spatial cut of the event horizon for model QS at affine
time $t= 2.5$.
In the absence of axisymmetry some of the circles on the torus which link the
hole  pinch off to zero faster than others -- the toroidal white
hole fissions into two white hole fragments. After the fissioning,
two ``pincers'' have formed between the sections, their tips are caustic
points.}
\label{fig:SP44}
\end{center}
\end{figure}

\begin{figure}[htb]
\begin{center}
 \epsfxsize=3.375in\leavevmode\epsfbox{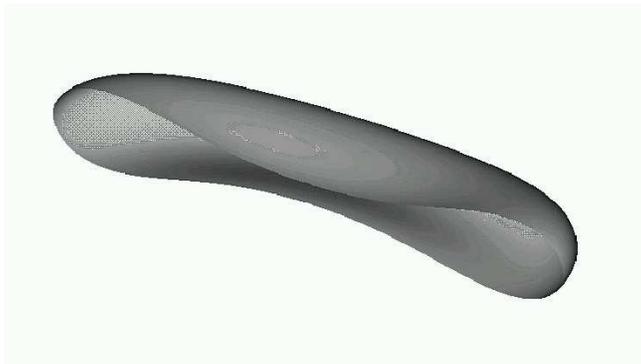}
\caption{Model BBH at an early time $t= 5.17$.
The surface corresponds qualitatively to Fig. \ref{fig:SP31} from the
quasispherical evolution but is now prolate instead of oblate.}
\label{fig:BBH20}
\end{center}
\end{figure}

\begin{figure}[htb]
\begin{center}
 \epsfxsize=3.375in\leavevmode\epsfbox{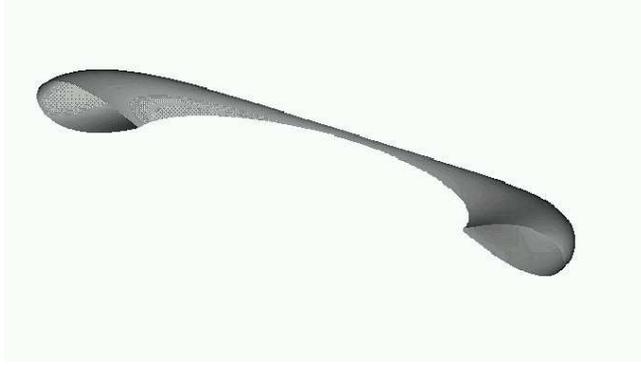}
\caption{Model BBH at time $t= 10.6$, when the event horizon is in its
toroidal phase.}
\label{fig:BBH40}
\end{center}
\end{figure}

\begin{figure}[htb]
\begin{center}
 \epsfxsize=3.375in\leavevmode\epsfbox{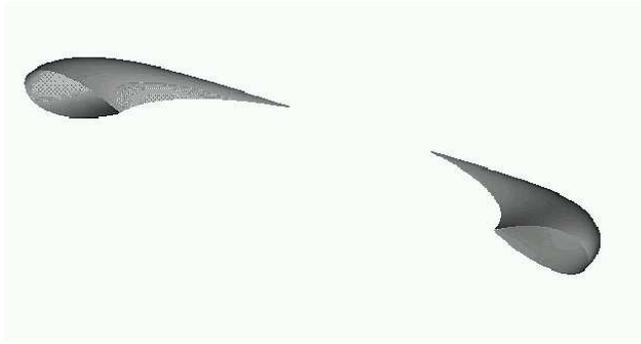}
\caption{Model BBH at time $t= 12.25$, after the event horizon has
fissioned into two components.}
\label{fig:BBH46}
\end{center}
\end{figure}

\begin{figure}[htb]
\begin{center}
 \epsfxsize=3.375in\leavevmode\epsfbox{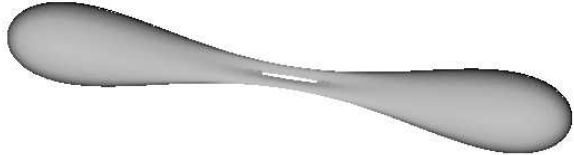}
\caption{Highly prolate model BBH2 at time $t=11.8$ showing the toroidal phase.
In the
axisymmetric prolate limit the hole in the torus shrinks to zero.}
\label{fig:2-BBH10}
\end{center}
\end{figure}

\begin{figure}[htb]
\begin{center}
 \epsfxsize=3.375in\leavevmode\epsfbox{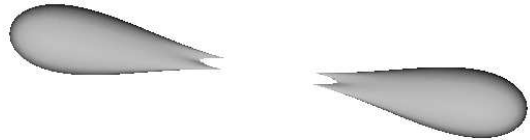}
\caption{Highly prolate model BBH2 at time $t=13.1$ showing two individual
white holes.
In the axisymmetric prolate limit the pincers merge into ``tips''
of the ``teardrops''.}
\label{fig:2-BBH11}
\end{center}
\end{figure}

In the following, we discuss our results for generic ellipsoids by the examples
of three individual evolutions. The models are called QS (quasispherical),
BBH and BBH2, the latter two corresponding to binary black hole merger
situations.
The parameters defining the models and the lifetimes $t_{end}$ from ${\cal
S}_{0}$
to the last point on the horizon (measured by the affine parameter $t$)
are given in Table 1. All the Figures \ref{fig:SP31}  --
\ref{fig:2-BBH11} showing snapshots of the evolutions have been constructed
using the method described in Sec. \ref{sec:tech}.
The images have been scaled to approximately the same size, and the snapshots
from the first two evolutions, Figs. \ref{fig:SP31} -- \ref{fig:BBH46}
show cut-away views to make the sharpness at the crossovers more
evident.

The simple heuristic picture of a generic situation suggested by the
oblate spheroidal collapse is indeed confirmed by generic models.
As an example we consider the quasi-spherical
ellipsoidal model QS with parameters given in Table I.
The time evolution of the resulting white hole horizon is depicted in the
three snapshots at constant $t$ shown in Figs. \ref{fig:SP31} --
\ref{fig:SP44}.
Figure \ref{fig:SP31} depicts the time $t= 1.78$. The surface shows obvious
oblateness and the deviation from axisymmetry is not yet apparent.
At this early stage, its history is similar to the oblate spheroidal case.
The surface is clearly bulged inwards, so the rays at the poles of the event
horizon cross first, and the event horizon becomes a torus, losing points at
an inner edge. This is shown at time $t=1.90$ in Fig. \ref{fig:SP34}, when the
deviation from axial symmetry is now evident. The circles on the torus which
link the hole shrink in size
at different rates and thus eventually pinch off to zero at different times.
When the first two circles pinch off, the toroidal white hole fissions into two
white hole fragments. After the fissioning, two ``pincers''
have formed between the sections. Being part of the boundary of $\cal X$,
their tips are caustic points and, in particular, the first caustic appears
just as the horizon is torn apart. The resulting two components are shown in
Fig. \ref{fig:SP44}, corresponding to time $t= 2.5$. At later
times, the individual components finally shrink to zero size at
$t= 2.88$.

If the reflection symmetry of the ellipsoidal model were broken, the torus
would
not pinch off {\em simultaneously} at two separate points. Instead, the first
meridian which shrank to zero would cause the topology to become spherical
again,
although with a sharp crease on its surface; and later, as a second meridian
shrank to zero, this sphere would fission
into two components. It also has to be expected, that in more general
situations one
can have intermediate stages where the white hole has more than one handle.
Within our
approach, such situations can indeed easily be generated by choosing
$\rho=const\times
\sigma$. In that case, The ${\cal S}_t$ slices develop an upward bulge as in
Fig. \ref{fig:slicings}
about each of the four umbilical points. Due to the reflection symmetry
the umbilical rays intersect in
pairs to form the first end points of the horizon, which results in
punching two holes into the corresponding white hole.

The results for our generic quasispherical model thus agree with the
expectations
suggested by the oblate spheroidal model. However, in order to model a black
hole merger
we require a geometry where the lifetimes of the individual components are
much larger than the apparent time of the merger. Indeed, such cases are
found easily by choosing a prolate ellipsoid, i.e. a model where one axis
is significantly more elongated than the other two, and taking the function
$\rho$ as
some constant.
Clearly, this is not a precise definition of prolateness, and there is no clear
transition
between the ``more prolate'' and ``more oblate'' ellipsoids.
Within the $\rho=const$ models, all of our results show the same qualitative
behavior of forming a torus, which then breaks apart into two pieces.
However, as one approaches the prolate spheroidal limit, the lifetimes of the
individual
components become infinite, while the relative size of the hole in the torus
shrinks to
zero, so that one recovers the earlier axisymmetric results. Models which are
sufficiently
close to a prolate spheroid thus show the (loosely defined) ``binary black hole
merger
behavior''.
The next two examples are chosen increasingly close to a prolate spheroid, and
show how a
binary-black-hole scenario emerges by deformation of the quasispherical model
while
the essential structural features -- the torus and the ``pincers'' -- remain.

Table I gives the parameters of the first example, called BBH.
Figure \ref{fig:BBH20} depicts an early time $t= 5.17$. The surface corresponds
qualitatively to Fig. \ref{fig:SP31} from the quasispherical evolution,
but is now prolate instead of oblate.
Again the surface is clearly bulged inwards. At the later time $t=10.6$
shown in Fig. \ref{fig:BBH40}, rays around the poles have already crossed
and the event horizon is in its toroidal phase, losing points at an inner
edge. Shortly afterwards, at $t= 12.25$, the toroidal white
hole has already fissioned into two white hole fragments, depicted in
Fig. \ref{fig:BBH46}. After the fissioning, two ``pincers''
have again formed between the sections. At time $t= 66.7$ the individual
components finally shrink to zero size.
Taking a cut of the null surface at $x=0$ produces the ``pair of pants''
depicted in Fig. \ref{fig:pop}. Again, this image was produced
with the method described in Sec. \ref{sec:tech}.

The last model, defined as BBH2 in Table I, is very close to
a prolate spheroid.
The appearance of the toroidal phase shown in Fig.
\ref{fig:2-BBH10} at $t= 11.8$ and the ``pincers'' shown in Fig.
\ref{fig:2-BBH11}
at $t= 13.1$ still remain, but both the hole and the pincers are much smaller
now.
Figures \ref{fig:2-BBH10} -- \ref{fig:2-BBH11} indicate how the limiting
transition to the
axisymmetric prolate case as the hole in the torus shrinks to zero
and the pincers shrink to the ``tips of teardrops''.
The two individual holes are long-lived and shrink to zero at $t= 644$.

Other models with $\rho$ approximately constant are found to share the same
features as
the models QS, BBH and BBH2: with increasing prolateness, the lifetimes of
the individual components increase. The average value
of $\rho$ mainly stretches the time scale, which increases with $\rho$.

\section{Discussion}
\label{sec:discussion}

In this paper we have introduced a class of solutions for the intrinsic
geometry of an event horizon. These solutions provide generic models
in the sense that the geometry does not possess any continuous
symmetries and the slicing dependent features will remain stable under
small perturbations of the slicing.
In contrast to the usual method of locating the
event horizon by tracing a null hypersurface (backwards for a black hole -
forward
for a white hole) in a numerically generated spacetime, the geometry here
is constructed as one part
of the initial data in a double-null evolution problem. The spacetime
thus is determined {\em a posteriori} by evolution of the full set of initial
data.

Our method is based on a conformal transformation from a null hypersurface in
Minkowski spacetime. The conformal factor is specified a priori in our
approach. In order to satisfy the projected Einstein vacuum equation,
the affine parameter is subject to an ODE along each ray, which has to be
integrated numerically in general. However, since all the numerics is reduced
to
this reparameterization problem, and all quantities have known explicit
dependence on the original affine parameter, our approach is essentially
analytic in nature.
This renders possible a simple and clear understanding of the
geometry and in particular the structure of caustics and crossovers.

The original aim of our approach is the construction of the double-null
initial data. The {\em intrinsic} geometry is given by Eq.
(\ref{eq:the_intrinsic_solution}). The {\em extrinsic} geometry will be studied
in a
forthcoming paper. A characteristic evolution code to evolve these data has
been constructed by the Pittsburgh numerical relativity group
\cite{cce,highp,wobb}  and evolutions
of these initial data are currently under investigation.

The study of the geometry of crossovers and caustics is particularly simple
in our case, because it is equivalent to the structure of the original
null hypersurface in Minkowski space. Numerical calculations are necessary only
to study slicing-dependent
features of the curved space geometry.

Apart from the geometry of the horizon,
it is also interesting to look for foliation dependent phenomena associated
with
the topology of the spatial slices.
Starting with Hawking's demonstration \cite{hawking} that the event horizon
of a stationary asymptotically flat spacetime has spherical topology, a number
of
results have appeared in the literature, which show that an event
horizon has to exhibit spherical topology on spacelike slices in the
late stages of black hole evolution
\cite{gannon,CW,gallw1,jacobson,browdygall,gallw2,siino1}.

It thus came as a surprise when a toroidal event horizon was found
in numerical simulations of the collapse of rotating clusters
\cite{torus}. In particular, this seemed to provide
the possibility of violating topological censorship \cite{fsw}, as was pointed
out by Jacobson and Venkatarami \cite{jacobson}.
However, because the crossover surface is spacelike all causal curves are
homotopically trivial, even though the slices of the
horizon contain handles \cite{toroid}. In effect, the hole in the torus
shrinks faster than the speed of light, in obedience to topological censorship.
Moreover, it is possible to choose slicings of the horizon which exhibit
arbitrarily
complicated topologies in the early phase of black hole
formation. Where these slices meet the spacelike crossover surface,
i.e. where new generators enter the horizon, they will in general not be
smooth. Some of the theorems which establish spherical
topology of the event horizon explicitly assume smoothness of the horizon, or
that no new generators enter.

The full interpretation of the results is difficult without a better
understanding
of collapse physics, i.e. without additional information from the construction
of the surrounding spacetime. These models thus serve a twofold purpose:
On the one hand, they are hoped to serve as a guide to explore new physical
situations -- in particular by setting up double-null initial data. On the
other hand, they make available a class of essentially analytic models that can
help to understand phenomena that appear in the study of 3D collapse with
complementary methods.
Examples are the spheroidal case, where these models help to understand
the results of previous numerical simulations. The availability of such
simple models is likely to be even more important in the more complicated
generic case.

In order to model a generic binary collision we modified the model used for
the axisymmetric collision only minimally; in particular, we retained the
choice
$\rho=const$. If instead the function $\rho$ has strong ray dependence,
in particular if we choose  $\rho= const \times \sigma$, a rich set of
histories
of topology change can be produced with our simple choice of slicing.
This fulfills the expectation that the
model is capable of yielding a large variety of black holes and can provide
the null data for studying a variety of waveforms.
However, in many of these models the lifetimes of the individual components
are of the order as the time scale for the appearance of nontrivial
topology. These cases should thus be interpreted as the formation of a single,
highly distorted black hole.

All models of the binary black hole merger type,
which displayed two horizon components with long individual lifetimes,
also exhibited an intermediate toroidal
phase before the final black hole formed with spherical topology. Since this
behavior is stable under perturbations,
toroidal event horizons turn out not to be a mere oddity but arise
generically in black hole collisions.

\begin{center}
{\bf ACKNOWLEDGMENTS}
\end{center}

\vspace{0.3cm}
This work has been supported by NSF PHY 9510895 to the University of
Pittsburgh. Computer time for this project has
been provided by the Pittsburgh Supercomputing Center. We are especially
grateful to Roberto G\'{o}mez and Luis Lehner for helpful discussions.

\begin{center}
\begin{table}
\begin{tabular}[htb]{l|c|c|c|c|c}
model & $\Xi$ & $\sqrt{\Xi^2-b^2}$ & $\sqrt{\Xi^2-a^2}$ & $\rho$
& $t_{end}$ \\
\hline
QS    & 1.0   & $0.99$             & $0.97$             & $\sigma_M/\sqrt{13}$
& 2.88      \\
BBH   & 1.0   & $0.66$             & $0.5 $             & $\sigma_M/\sqrt{13}$
& 66.7      \\
BBH2  & 1.0   & $0.66$             & $0.64$             & $\sigma_M/\sqrt{13}$
& 644
\end{tabular}
 \label{table:parameters}
 \caption{Parameters defining the generic models studied in detail;
$t_{end}$ is the lifetime from ${\cal S}_{0}$
to the last point on the horizon as measured by the affine parameter $t$.}
\end{table}
\end{center}
%
%
\end{document}